\documentclass[aps,pra,showpacs,showkeys,superscriptaddress,twocolumn]{revtex4-1}
\usepackage[colorlinks=true,linkcolor=blue,anchorcolor=blue,citecolor=blue,urlcolor=blue,breaklinks=true]{hyperref}
\usepackage{graphicx}% Include figure files
\usepackage{graphics}
\usepackage{amsmath}
\allowdisplaybreaks[4]
\usepackage{amssymb}
\usepackage{slashed}
\usepackage{latexsym}
\usepackage{epsfig}
\usepackage{amsbsy}
\usepackage{array}
\usepackage{amssymb}
\usepackage{setspace}
\usepackage{bm}
\usepackage{lipsum}
\usepackage{mathrsfs}
\usepackage{float}
\usepackage{color}
\usepackage[T1]{fontenc}
\usepackage{mathptmx}
\usepackage{ulem}

\allowdisplaybreaks[2]
\DeclareMathAlphabet{\mathcal}{OMS}{cmsy}{m}{n}
\DeclareSymbolFont{largesymbols}{OMX}{cmex}{m}{n}

\begin{document}
\address{Department of Physics, Nanjing University, Nanjing 210093, China}
\address{School of Physics and Electronic Engineering, Linyi University, Linyi 276000,China}
\address{Nanjing Institute of Proton Source Technology, Nanjing 210046, China}
\author{Hao Zhao$^{1, 2}$}
\author{Yong-Long Wang$^{2, 3}$}
 \email{wangyonglong@lyu.edu.cn}
\author{Cheng-Zhi Ye$^{2}$}
\author{Run Cheng$^{1, 2}$}
\author{Guo-Hua Liang$^{1}$}
\author{Hui Liu$^{1  }$}
\title{Quantum mechanics of fermion confined to a curved surface in Foldy-Wouthuysen representation}

\begin{abstract}
In Foldy-Wouthuysen representation, we deduce the effective quantum mechanics for a particle confined to a curved surface by using the thin-layer quantization scheme. We find that the spin effect caused by confined potential as the results of relativistic correction in the non-relativistic limit. Furthermore, the spin connection appeared in curved surface which depends on curvature contributes a Zeeman-like gap in the relativistic correction term. In addition, the confined potential also induces a curvature-independent energy shift, which is from the zitterbewegung effect. As an example, we apply the effective Hamiltonian to torus surface, in which we obtain expectantly the spin effects related to confined potential. Those results directly demonstrate the scaling of the uncommutation of the non-relativistic limit and the thin-layer quantization formalism.

\end{abstract}
\maketitle
\section{Introduction}
\indent With the rapid development of nano-technology, more and more nanostructures can be fabricated with complex geometrical and topological structures\cite{Ebbesen1992Large, Chopra1995Boron, Dmitry2009Longitudinal, Jiao2009Narrow}. The presence of geometrical and topological deformations leads those deformed nanosystems with various novel quantum properties. Specifically, the curvature effects were widely investigated in thin magnetic shells\cite{PhysRevLett.112.257203, meng2014elastic}, nematic shells\cite{PhysRevLett.108.207803}, titania single crystals~\cite{yang2014titania}, smectic liquid crystals~\cite{ncom10236}, quantum spin Hall system~\cite{huang2017bending}, photonic crystal fiber~\cite{Beravate1601421}, domain wall pinning~\cite{PhysRevB.92.104412}, domain wall motion~\cite{PhysRevB.93.094418}, antiferromagnets~\cite{pylypovskyi2020curvilinear} and so on. For two-dimensional (2D) curved surfaces, the thin-layer quantization scheme~\cite{AOP.63.586, PhysRevA.23.1982, Wang2016Quantum} is an effective and suitable method, which has been successfully employed to deduce effective Maxwell's equation\cite{PhysRevA.78.043821, PhysRevA.97.033843, PhysRevA.100.033825},
effective Schr\"{o}dinger equation\cite{PhysRevLett.100.230403}, effective Pauli equation\cite{JPSJ.80.073602, PLA.380.2876, PhysRevA.90.042117} and effective Dirac equation\cite{PLA.380.3036, PhysRevA.98.062112, PhysRevA.101.053632, AOP.323.2044, AOP.275.297, MPLA.27.1250016, PhysRevA.48.1861} for electromagnetic field and electrons confined to a 2D curved surface. There are two important results, geometric potential~\cite{AOP.63.586, PhysRevA.23.1982} and geometric momentum~\cite{Liu2011Geometric, Wang2017Geometric}, which have been proved experimentally in topological crystal\cite{PhysRevLett.104.150403} and plasmon polarization\cite{Guided2015OE}, respectively.\\

The thin-layer quantization formalism is valid and successful, but there are still some difficulties that have to be reconsidered, such as the uncommutation of the non-relativistic limit and the thin-layer quantization process~\cite{Wang2021}. Specifically, for relativistic fermions in the external field, the non-relativistic limit contributes a magnetic moment remained in the effective Pauli equation, surprisingly without the effective spin-orbit coupling and corresponding energy correction induced by external field. In superconducting materials and valley electronics devices, the effects of field-induced Darwin term\cite{PhysRevB.59.7155,PhysRevB.63.052503} and spin precession\cite{PhysRevB.84.195463,PhysRevB.86.165411,PhysRevB.86.045456} have been considered. In order to describe these two effects in the non-relativistic limit, we will reconsider the Dirac equation in curved spacetime by introducing the Foldy-Wouthuysen representation (FWR)\cite{PhysRev.78.29,PhysRev.111.1011,JMP.44.2952,JMP.50.122302,PhysRevA.77.012116}.

In this paper, in the spirit of the thin-layer quantization scheme we will discuss the effective dynamics of Dirac fermion confined to a curved surface in FWR. This paper is organized as follows. In Sec. II, we briefly review the Dirac equation in curved spacetime and the Foldy-Wouthuysen transformation (FWT). In Sec. III, in the presence of an external field, we first consider a Dirac fermion confined to a curved surface $S$, and then we deduce the corresponding effective Hamiltonian in FWR by using the thin-layer quantization approach. Interestingly, there appears a Zeeman effect induced by the coupling of spin connection and the external field. As an example, a torus surface is considered, in which the curvature-induced Zeeman splitting is demonstrated. Finally, we present our conclusions in Sec. IV.
\section{Dirac equation and its Foldy-Wouthuysen representation}
In this section, we will briefly review the Dirac equation in a curved spacetime and FWT in the presence of an external field. A relativistic electron in a curved space-time can be described by the Dirac equation, that is
\begin{eqnarray}\label{eq}
% \nonumber to remove numbering (before each equation)
(i {\gamma}^\mu \nabla_\mu-{m})\Phi=0,
\end{eqnarray}
where $\nabla_\mu$ denotes a covariant derivative with $\nabla_\mu=\partial_\mu+\Gamma_\mu$, wherein $\Gamma_\mu$ is a spin connection and $\mu=0, 1, 2, 3$ describe the four coordinate variables of a curved space-time, and ${\gamma}^\mu$ stands for a Dirac matrix in a curved space-time, which can be expressed as
\begin{align*}
  {\gamma}^\mu={E^\mu}_\alpha {\gamma}^\alpha,
\end{align*}
here $\gamma^{\alpha}$ stands for a Dirac matrix in a flat space-time, wherein $\alpha=0, 1, 2, 3$ describes the four coordinate variables of flat space-time, and ${E^\mu}_\alpha$ are vierbeins that can be defined by
\begin{align*}
  {E^\mu}_\alpha=\frac{\partial q^\mu}{\partial x^\alpha},
\end{align*}
where $q^\mu$  stands for a coordinate variable of a curved space-time, and $x^\alpha$ stands for a coordinate variable of a flat space-time. With the vierbeins ${E^{\mu}}_{\alpha}$, the inverse of metric tensor defined in a curved space-time $G^{\mu \nu}$ can be given by
\begin{equation*}
  G^{\mu \nu}={E^\mu}_\alpha {E^\nu}_\beta \eta^{\alpha \beta}
\end{equation*}
where $\eta^{\alpha \beta}$ is the inverse metric tensor in a flat space-time with $\eta^{\alpha \beta}=diag(-1,1,1,1)$. And the spin connection $\Gamma_\mu$ can be expressed as
\begin{equation*}
  \Gamma_\mu=\frac{1}{4} {E_{\nu}}^{\alpha}(\partial_\mu E^{\nu\beta}+{\Gamma^\nu}_{\kappa\mu} E^{\kappa\beta}) \Sigma_{\alpha \beta},
\end{equation*}
with ${\Gamma^\nu}_{\kappa\mu}$ being the Christoffel symbol and $\Sigma_{\alpha \beta}=[\gamma_\alpha, \gamma_\beta]/2$.
In the present paper, the space is curved and the time is taken as a constant, and thus $ds^2$ can be described as
\begin{equation*}
  ds^2=-G_{00}dq^0 dq^0+ G_{AB}dq^A dq^B,
\end{equation*}
where $G_{00}=1$ and $A, B=1, 2, 3$ stand for the three coordinate variables of a curved space. Therefore, ${E_\nu}^\alpha$ can be simplified as
\begin{equation*}
  E^\alpha_\nu=\left(
                 \begin{array}{cc}
                   1 & 0 \\
                   0 & {E_A}^I \\
                 \end{array}
               \right),
\end{equation*}
where $I=1, 2, 3$ stand for the three coordinate variables of a flat space. And the space components of Dirac matrices $\gamma^A$ equal to ${E^A}_I \gamma^I$. By introducing a Dirac matrix $\beta=\gamma^0$, the Dirac equation can be rewritten as
\begin{equation*}
  i\frac{\partial\Phi}{\partial t}=H\Phi,\  \
\end{equation*}
where $H$ is a Dirac Hamiltonian as
\begin{equation*}
H=-i\beta {\gamma}^A \nabla_A+\beta m.
\end{equation*}
In view of the uncommutation of non-relativistic limit and thin-layer quantization scheme, the influence of non-relativistic limit on the geometric effects, which are obtained in the thin-layer quantization formalism, is worthy of discussion further. Without loss of generality, we boldly assume Dirac fermion in an external electric field and specifically discuss in FWT. In the presence of external field, Dirac fermion can be described by a Dirac Hamiltonian, that is
\begin{equation}\label{eq11}
H=-i{\gamma}^A \nabla_A+ m + V,
\end{equation}
where the the $\nabla_A=\partial_A+\Gamma_A+A_A$, within $A_A$ is a gauge potential such as magnetic vector potential, $V$ is the potential of Dirac fermion in an external field.
By introducing the matrix $\beta$, the Dirac Hamiltonian Eq.~\eqref{eq11} can be reepressed as
\begin{eqnarray}\label{eq12}
% \nonumber to remove numbering (before each equation)
H= -i\beta{\gamma}^A \nabla_A +\beta m+ \beta V,
\end{eqnarray}
With FWT, the Dirac Hamiltonian Eq.~\eqref{eq11} can be simplified in FWR as
\begin{eqnarray}
% \nonumber to remove numbering (before each equation)
H=\beta m + \mathcal{E} + \mathcal{O},
\end{eqnarray}
where $\mathcal{E}=\beta V$ is even form that commutes with $\beta$, and $\mathcal{O}= -i\beta{\gamma}^A \nabla_A$ is odd form that anti-commutes with $\beta$. As $S$ does not depend on time with $S=\frac{-i\beta \mathcal{O}}{2m}$, the first unitary operator can be directly described by $e^{iS}$, we can obtain a new Dirac Hamiltonian as
\begin{eqnarray}\label{eqfw}
\begin{split}
% \nonumber to remove numbering (before each equation)
H_{FW}=e^{iS}H e^{-iS}=&\beta m+\beta \mathcal{E}+\frac{\beta}{2m}\mathcal{O}^2-\frac{1}{8m^2}[\mathcal{O}, [\mathcal{O}, \mathcal{E}]]\\
&+\frac{\beta}{2m}[\mathcal{O}, \mathcal{E}]-\frac{1}{3m^2}\mathcal{O}^3+...,
 \end{split}
\end{eqnarray}
where it is worthwhile to notice that the even part $\mathcal{E}$ is invariant, but the odd part $\mathcal{O}$ is expanded into a new even part and an odd part. The final goal of FWT is to transform the original Dirac Hamiltonian into a block diagonal one, in other words, to eliminate the off-diagonal part that is odd part. In terms of the odd part in Eq.~\eqref{eqfw}, the second unitary operator can be expressed as $e^{iS_2}$ with $S_2=\frac{-i\beta \mathcal{O}_2}{2m}$ and $\mathcal{O}_2=\frac{\beta}{2m}[\mathcal{O}, \mathcal{E}]$. For the rest odd part, the third unitary operator can be given by $e^{iS_3}$, with $S_3=\frac{-i\beta \mathcal{O}_3}{2m}$ and $\mathcal{O}_3=-\frac{1}{3m^2}\mathcal{O}^3$. It is clear that the transformation only converges if the $m$ is much larger than the others in Eq.~\ref{eq11}. In the case of the non-relativistic limit, the first four terms of $H_{FW}$ are a valid approximation due to the rapid decay of the power series of $\frac{1}{m}$, and we will consider only them next in FWR.
\section{Foldy-Wouthuysen representation of Dirac equation on curved surface}
\indent In this section, the effective quantum mechanics will be discussed in FWR for the Dirac fermion confined to a 2D curved surface. Without loss of generality, the 2D curved surface $S$ is originally embedded in a three-dimensional Euclidean space. For the sake of convenience, we can adapt a curvilinear coordinate system with $(q_1, q_2, q_3)$, where $q_1$ and $q_2$ are two tangent coordinate variables of $S$ and $q_3$ is a normal one. Therefore, in the small neighborhood of $S$ denoted as $\Xi S$, the position of point can be parametrized by
\begin{equation}\label{eqr}
  \vec{R}(q_1, q_2, q_3)=\vec{r}(q_1, q_2)+q_3\vec{n}(q_1, q_2)
\end{equation}
where $\vec{n}(q_1, q_2)$ is the normal unit basis of $S$, which is a function of
$q_1$ and $q_2$. With Eq.~\ref{eqr}, the covariant elements of the metric tensor $G_{AB}$ in $\Xi S$ can be defined as
\begin{equation*}
  G_{AB}=\frac{\partial \vec{R}}{\partial q^A}\cdot\frac{\partial \vec{R}}{\partial q^B}.
\end{equation*}
And on  $S$ described by $\vec{r}(q_1, q_2)$, the covariant elements of the metric tensor $g_{AB}$ can be given by
\begin{equation*}
  g_{ab}=\frac{\partial \vec{r}}{\partial q^a}\cdot\frac{\partial \vec{r}}{\partial q^b},
\end{equation*}
where $a, b=1, 2$ denote the two tangent coordinate variables of $S$. According to the two definitions of $G_{AB}$ and $g_{ab}$, it is easy to prove that the relationship between $G_{ab}$ and $g_{ab}$ satisfies the following equation
\begin{equation}\label{eq2}
\begin{split}
  &G_{ab}=g_{ab} +q_3 [\alpha g+(\alpha g)^T]_{ab}+q^2_3 (\alpha g\alpha^T)_{ab},
\end{split}
\end{equation}
and $G_{a3}=G_{3a}=0$, $G_{33}=1$,  where $\alpha$ is the Weingarten curvature tensor that can be expressed as
\begin{equation*}
  \alpha_{ab}=\frac{\partial \vec{r}}{\partial q^a}\cdot\frac{\partial \vec{n}}{\partial q^b}.
\end{equation*}
In $\Xi S$ and with the adapted coordinate system of $(q_1, q_2, q_3)$, the vierbeins can be expressed as
\begin{equation}
  E_A^I=\left(
          \begin{array}{cc}
            E_a^i & 0 \\
            0 & 1 \\
          \end{array}
        \right)
\end{equation}
where
\begin{equation}\label{eq4}
\begin{split}
  &E_a^i=e_a^i+q_3\alpha^b_a e^i_b,\\
  &E_i^a=e_i^a-q_3\alpha^b_a e^b_i+O[q^2_3].
\end{split}
\end{equation}
With these vierbeins and their inverses, the reduced Dirac matrices $\gamma^a$ can be expressed as $\gamma^a=E^a_i\gamma^i$ and $\gamma^3$ is invariant.\\
\subsection{Effective Hamiltonian on surface in FWR}
In the spirit of the thin-layer quantization scheme, we can introduce a confining potential $V(q_3)$ to confine Dirac fermion to the curved surface $S$. For simplicity, the confining potential can be taken as a square well that vanishes on $S$ and infinity off $S$. Therefore, the Dirac Hamiltonian should be replaced by
\begin{eqnarray}\label{eq1}
% \nonumber to remove numbering (before each equation)
H=-i \beta {\gamma}^A \nabla_A+{\beta m}+{\beta V(q_3)},
\end{eqnarray}
where $\nabla_A=\partial_A+\Gamma_A$, the term $-i \beta {\gamma}^A \nabla_A$ is odd. Therefore, the first unitary operator for FWT can be given by $U_{1}=e^{iS_1}$, here $S_1=\frac{-{\gamma}^A \nabla_A}{2m}$. The first transformed Hamiltonian can be expressed as
\begin{eqnarray}
\begin{split}
% \nonumber to remove numbering (before each equation)
H_{FW}=&e^{iS_1}H e^{-iS_1}\\
=&\beta m+\beta V(q_3)+\frac{\beta}{2m}({{\gamma}^A \nabla_A})^2\\
&+\frac{1}{8m^2}[\beta {\gamma}^A \nabla_A, [\beta {\gamma}^A \nabla_A, \beta V(q_3)]]\\
&+\frac{-i\beta}{2m}[\beta {\gamma}^A \nabla_A, \beta V(q_3)]-\frac{i}{3m^2}({ \beta {\gamma}^A \nabla_A})^3+...,
\end{split}
\end{eqnarray}
in the right hand side, the former four terms are block and even, but the latter two terms are odd, which can provide the second unitary operator $U_2=e^{iS_2}$ and the third unitary operator $U_3=e^{iS_3}$, here $S_2=\frac{-1}{4m^2}[\beta {\gamma}^A \nabla_A, \beta V(q_3)]$, $S_3=\frac{-\beta}{6m^3}({\beta {\gamma}^A \nabla_A})^3$. Transformed by $e^{iS_2}$ and $e^{iS_3}$, the new Hamiltonian can be described by
\begin{eqnarray}\label{eq10}
\begin{split}
H^1_{FW}=&e^{iS_3}e^{iS_2}H_{FW} e^{-iS_2}e^{-iS_3}\\
=&e^{iS_3}e^{iS_2}e^{iS_1}H e^{-iS_1}e^{-iS_2}e^{iS_3}\\
=&\beta m+\beta V(q_3)+\frac{\beta}{2m}({{\gamma}^A \nabla_A})^2\\
&+\frac{1}{8m^2}[\beta {\gamma}^A \nabla_A, [\beta {\gamma}^A \nabla_A, \beta V(q_3)]]+....
\end{split}
\end{eqnarray}
Obviously, $H^1_{FW}$ is a power series of $\frac{1}{m}$, which would converge when $m$ is dominant in Hamiltonian. In other words, the Hamiltonian is effect in non-relativistic limit. Through three transformations, in Eq.~\eqref{eq10} the former four terms are very greater than the rest terms for a certain mass. Without loss of generality, in the present paper the former four terms is taken as the Dirac Hamiltonian denoted as $H_{EFW}$.

\indent According to the thin-layer quantization approach, we can confine a fermion to $S$ by introducing a confining potential $V(q_3)$. In the squeezing process, the differential homeomorphism transformation induced by curvature have to be considered. Specifically, the curvature-induced transformation can be described by the relationship, $G = f^2 g$, where $G$ is the determinant of the metric tensor matrix $G_{AB}$ {defined in $\Xi S$}, $g$ is that of $g_{ab}$ on $S$, and $f$ is the rescaling factor with $f=1+Tr(\alpha)q_3+det(\alpha)q_3^2$. Under the rescaling transformation, the relationship between an introduced new wave function $\psi$ and the initial wave function $\Phi$ can be expressed in the following form
\begin{equation}\label{eq6}
  \psi=f^{1/2}\Phi,
\end{equation}
and the transformed Dirac Hamiltonian satisfies the following transformation
\begin{equation}
\bar{H}_{FW}\rightarrow f^{\frac{1}{2}}H_{FW}f^{-\frac{1}{2}}.
\end{equation}
And then we can expand the Hamiltonian $\bar{H}_{FWT}$ as a power series of $q_3$ and take the limit $q_3=0$, the effective Dirac Hamiltonian confined to $S$ can be obtained as
\begin{equation}
  \bar{H}_{EFW}=\beta m+\beta V(q_3)+ H^{\prime}_{FW}+ H^{\prime\prime}_{FW}
\end{equation}
where
\begin{equation}\label{eq7}
\begin{split}
  H^{\prime}_{FW}=&\frac{\beta}{2m}\{[\bar{\gamma^a} (\partial_a+\Omega_a)][\bar{\gamma^b} (\partial_b+\Omega_b)]+\partial_3^2\\
  &-\gamma^3\alpha_b^a e_i^b \gamma^i\partial_a+\gamma^3\alpha_b^a e_i^b \gamma^i (\Omega_a+\frac{1}{2}\epsilon^a_b\bar{\gamma_a}\gamma^3\alpha_a^a)\\ &+\gamma^3\bar{\gamma^a}\partial_3\Gamma_a+\partial_3\Gamma_3+|\epsilon^a_b|\alpha^a_b\alpha^b_a/2\},
\end{split}
\end{equation}
wherein $\bar{\gamma^a}=e^a_{i}\gamma^i$ is the reduced Dirac matrix defined on $S$, $\Omega_a$ is the the normal part of spin connection $\Gamma_a$, which does not depend on $q_3$. And $\epsilon$ is a second order matrix whose anti-diagonal elements are one and the others are zero. And $H^{\prime\prime}_{FW}$ is
\begin{equation}\label{Eq17}
 \begin{split}
 H^{\prime\prime}_{FW}=&-\frac{V(q_3)}{m}H^{\prime}_{FW}
  -\frac{\beta}{8m^2}[2\gamma^3(\partial_3 V(q_3))\bar{\gamma^a} (\partial_a+\Omega_a)\\
  &+4(\partial_3 V(q_3))\partial_3+(\partial_3^2V(q_3))].
 \end{split}
\end{equation}
In the above discussion, $H^{\prime}_{FW}$ is the effective result of $\frac{\beta}{2m}({{\gamma}^A \nabla_A})^2$ in $\bar{H}_{EFW}$ by using the thin-layer quantization approach, and $H^{\prime\prime}_{FW}$ is that of $\frac{1}{8m^2}[\beta {\gamma}^A \nabla_A, [\beta {\gamma}^A \nabla_A, \beta V(q_3)]]$. $H^{\prime}_{FW}$ and $H^{\prime\prime}_{FW}$ are both block diagonal Hamiltonians which are proportional to $\beta$, so the two block in them are opposites of each other.\\

\indent In the thin-layer quantization formalism, the last step is to separate analytically the surface component from the normal one for $\bar{H}_{EFW}$. It is easy to accomplish for the $\beta m+\beta V(q_3)+H^{\prime}_{FW}$, but it is invalid for $H^{\prime\prime}_{FW}$ that results from the coupling of $V(q_3)$ and the tangent momentum. Fortunately, the magnitude of $H^{\prime\prime}_{FW}$ is very smaller than other terms, and it can be simply taken as perturbation. In $\beta m+\beta V(q_3)+H^{\prime}_{FW}$, the normal Hamiltonian is
\begin{equation}
 \begin{split}
 H_{n}=\beta V(q_3)+\beta\partial^2_3
 \end{split}
\end{equation}
and the surface effective Hamiltonian is
\begin{equation}\label{eq12}
\begin{split}
  H_s=&\beta m+\frac{\beta}{2m}\{[\bar{\gamma^a} (\partial_a+\Omega_a)][\bar{\gamma^b} (\partial_b+\Omega_b)]+\\
  &-\gamma^3\alpha_b^a e_i^b \gamma^i\partial_a+\gamma^3\alpha_b^a e_i^b \gamma^i (\Omega_a+\frac{1}{2}\epsilon^a_b\bar{\gamma_a}\gamma^3\alpha_a^a)\\ &+\gamma^3\bar{\gamma^a}\partial_3\Gamma_a+\partial_3\Gamma_3+|\epsilon^a_b|\alpha^a_b\alpha^b_a/2\}.
\end{split}
\end{equation}

\indent As a block diagonal Hamiltonian, the two block of $H_s$ are opposite each other, the positive energy block corresponds to Pauli equation and the other does to the negative state. In $H_s$, $\Omega_a$ plays the role of pseudo-magnetic field that is given in Ref.~\cite{PhysRevA.98.062112}, which is induced by Gauss curvature. And the spin-orbit coupling term $-\gamma^3\alpha_b^a e_i^b \gamma^i\partial_a$ is in agreement with that in Ref.~\cite{PhysRevA.98.062112}. The rest terms $\gamma^3\alpha_b^a e_i^b \gamma^i (\Omega_a+\frac{1}{2}\epsilon^a_b\bar{\gamma_a}\gamma^3\alpha_a^a) +\gamma^3\bar{\gamma^a}\partial_3\Gamma_a$ play the role of Zeeman-like splitting that is contributed by the gradient of vierbein fields and the normal spin connection together. In the presence of curvature, the SOC is deformed. As a result, there is the Zeeman-like splitting in the considered system. The last term $|\epsilon^a_b|\alpha^a_b\alpha^b_a/2$ is the known geometric potential~\cite{PhysRevA.23.1982} that depends on the anti-diagonal elements of Wengarten matrix.

\subsection{Effect of confined potential}
\indent In comparison with the Pauli equation, the non-relativistic limit of Dirac equation, the effective Dirac equation contains additional high power terms of $\frac{1}{m}$. It is straightforward that $H^{\prime\prime}_{FW}$ is a function of $\frac{1}{m^2}$, which is a relativistic effect added by the presence of external field. When a fermion is moving in an external field, which will feel a force given by the external field. In other words, the momentum is coupled with the spin in Dirac equation, and the coupling can be modified by the external field. As a result, the spin precession will occur in the external field. In Eq.~\eqref{eq1}, the confining potential plays a same role of mass. It is easy to check that in the non-relativistic limit the confining potential can provide a mass correction that is proportional to $\frac{V(q_3)}{m}$. More importantly, the confining potential $V(q_3)$ is a function of $q_3$ not a constant as $m$, the normal gradient $\partial_3V$ can provide a force to modify the fermion by deforming SOC. It is worthwhile to notice that on a curved surface the curvature will couple with the spin of Dirac fermion that will deform the original SOC. The deformation can induce a Zeeman-like splitting.
\indent Furthermore, the second-order $\partial^2_3V(q_3)$ can only fluctuate the normal position of fermion that is caused by the interference between positive energy state and the negative energy one, which is called zitterbewegung. As a result, the zitterbewegung will contribute a constant energy shift to the fermion energy level.\\

\indent To illustrate the actions of confining potential on the effective Dirac dynamics in FWR, we consider three simple examples for the confining potential as: (a) $V(q_3)=m\omega |q_3|$; (b) $V(q_3)=m\omega q^2_3$ with $\omega$ is a constant; (c) a deep square well with a certain width $L$. In the case of (a), the $\partial_3 V(q_3)$ is a constant that depends on the frequency $\omega$, which will induce a certain spin precession correction and Zeeman-like splitting, no zitterbewegung for vanishing $\partial^2_3 V(q_3)$. In the spirit of the thin-layer quantization scheme, the effective Dirac Hamiltonian should be replaced by the following form
\begin{equation}
  H_{eff}=H_s-\frac{\beta}{4m}\gamma^3\omega\bar{\gamma^a} (\partial_a+\Omega_a)
\end{equation}
and the normal Hamiltonian would be corrected to
\begin{equation}
 \begin{split}
 H_{n}=\beta V(q_3)+\beta\partial^2_3-\frac{\beta}{2m}\omega\partial_3
 \end{split}
\end{equation}
where the $q_3$-dependent terms are ignored. In the case of (b), $\partial_3 V(q_3)=2m\omega q_3$ can not significantly affect the spin precession and Zeeman-like splitting, because of the dependence of $q_3$. In contrast to case (a), the difference is the zitterbewegung in the case of (b). Therefore, the effective Dirac Hamiltonian should be replaced by
 \begin{equation}
  H_{eff}=H_s-\frac{\beta}{4m}\omega.
\end{equation}
In the case of (c), $H^{\prime\prime}_{FW}$ vanishes on $S$, the confined potential does not affect the effective Dirac Hamiltonian.

\indent According to the above discussions, what is considerably different from the known conclusion given by R. C. T. de Costa~\cite{PhysRevA.23.1982} is that the details of confining potential does not affect the effective quantum dynamics on a curved surface. In FWR, the specific form of confining potential plays an important role in the effective Dirac Hamiltonian. Specifically, the effects of the details of confining potential can measure the scaling of uncommutation of the non-relativistic limit and the thin-layer quantization process.

\subsection{An example: torus surface}

\begin{figure}
  \centering
  \includegraphics[width=5.0cm]{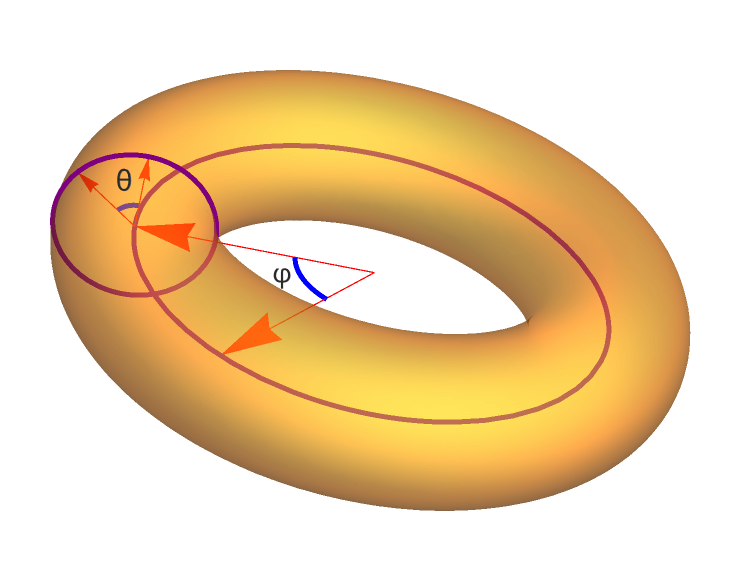}
  \caption{A torus surface with curvilinear coordinate {variables $\theta$ and $\varphi$.} }\label{fig1}
\end{figure}

In this section, the previous discussions will be applied to a torus that can be parametrized by $(\theta,\varphi)$ shown in Fig.~\ref{fig1}. Therefore, the effective Hamiltonian in Eq.~\eqref{eq12} is
\begin{equation}
\begin{split}
  H_s=&\beta m+\frac{\beta}{2m}\{[\bar{\gamma^\theta}\partial_\theta+\bar{\gamma^\varphi} (\partial_\varphi+i\frac{\sin\theta}{2}\Sigma_3)]^2\\
  &-\gamma^3(\frac{1}{r}\bar{\gamma^
  \theta}\partial_\theta+\frac{\cos\theta}{(R+r\cos\theta)}\bar{\gamma^\varphi}\partial_\varphi)\\
  &-i\frac{\cos\theta}{(R+r\cos\theta)}\frac{\sin\theta}{2}\Sigma^\varphi,
\end{split}
\end{equation}
and the contributions given by the confining potential is
\begin{equation}
\begin{split}
&H^{\prime\prime}_{FW}=-\frac{V(q_3)}{m}H^{\prime}_{FW}
-\frac{\beta}{8m^2}\{2\gamma^3(\partial_3 V(q_3))[\bar{\gamma^\theta}\partial_\theta\\
&+\bar{\gamma^\varphi} (\partial_\varphi+i\frac{\sin\theta}{2}\Sigma_3)]+4(\partial_3 V(q_3))\partial_3+(\partial_3^2V(q_3))\},
\end{split}
\end{equation}
where $\bar{\gamma^\theta}=e^\theta_i\gamma^i$ and $\bar{\gamma^\theta}=e^\varphi_i\gamma^i$ are tangent Dirac matrices, $\Sigma_3=i\gamma_1\gamma_2$ is the normal one, and $\Sigma^\varphi$ is the $\varphi$ Dirac matrix.

\indent On the torus surface, the spin connection contained in the operator $\nabla_\theta$ vanishes, and that in $\nabla_{\varphi}$ is $\frac{i\sin\theta\Sigma_3}{2}$. The novanishing term plays a role of pseudo-magnetic field that leads to the appearance of Zeeman-like gap depending on position. In addition, in $H^{\prime\prime}_{FW}$ the couplings depending on curvature can enlarge the Zeeman-like gap in the $\varphi$ direction. When $\theta=0$ or $\pi$, the spin connection vanishes, and the Zeeman-like gap also vanishes. And when $\theta=\frac{\pi}{2}$ or $\frac{3\pi}{2}$, the coefficient of Zeeman-like gap, $\frac{\cos\theta}{(R+r\cos\theta)}$, will vanishes, the spin gap in $H_s$ will vanish, too.  Therefore, we can experimentally observe the spin degeneracy of particles that is determined by the confining potential.
\indent The spin-orbit interaction terms in $H_s$ depend on $\gamma^\theta\partial_\theta$ and $\bar{\gamma^\varphi}\partial_\varphi$ by the coefficients being functions of curvature. In other words, the Zeeman-like splitting can be controlled by modifying the curvature of curved surface. Due to the Wengarten matrix $\alpha$ is diagonal, the known geometric potential vanishes on the torus surface.

\section{CONCLUSION}
In this paper, we have briefly reviewed the Dirac equation describing the relativistic electron confined to a curved surface, and the Foldy-Wouthuysen transformation. In FWR, by using the thin-layer quantization approach we have deduced the effective Dirac Hamiltonian for the relativistic particle confined to a curved surface. Strikingly, there are two contributions in the final effective Hamiltonian, which are provided by the introduced confining potential. As the results of FWT, the two contributions are spin-orbit couple and zittebewegung, respectively. Furthermore, the spin connection determined by the geometry of curved surface is coupled with spin, the coupling contributes a Zeeman-like splitting. These results directly demonstrate the uncommutation of the non-relativistic limit and the thin-layer quantization process which can be employed to quantum measure.\\

\indent The above discussions have been applied to a torus surface, which are valid. The spin connection plays the role of pseudo-magnetic field, and couples with spin as a SOC, which can induce the Zeeman-like gap. These results can be employed to precisely control the particular properties of spintronic devices and valley electronics by designing their geometries.

\section{ACKNOWLEDGEMENTS}
This work is jointly supported by the Natural Science Foundation of Shandong Province of China (Grant No. ZR2020MA091), the National Major state Basic Research and Development of China (Grant No. 2016YFE0129300), the National Natural Science Foundation of China (Grants No. 11690030, No. 11690033, No. 11535005, No. 61425018).
\bibliography{reference}
\end{document}